\documentclass[longauth]{aa} % for the letters 
\pdfoutput=1
\usepackage{graphicx}
\usepackage{txfonts}
\usepackage{natbib}
\usepackage[colorlinks=true, citecolor=blue]{hyperref}
\bibpunct{(}{)}{;}{a}{}{,}

\begin{document} 

\title{Discovery of a nitrogen-enhanced mildly metal-poor binary system: Possible evidence for pollution from an extinct AGB Star}

\author{
Jos\'e G. Fern\'andez-Trincado\inst{1,2,3},
Ronald Mennickent\inst{3},
Mauricio Cabezas\inst{4}, 
Olga Zamora\inst{5,6},
Sarah L. Martell\inst{7,8},
Timothy C. Beers\inst{9},
Vinicius M. Placco\inst{9},
David M. Nataf\inst{10},
Szabolcs~M{\'e}sz{\'a}ros\inst{11,12},
Dante Minniti\inst{13, 14, 15},
Dominik R. G. Schleicher\inst{3},
Baitian Tang\inst{16},
Angeles P\'erez-Villegas\inst{17},
Annie C. Robin\inst{2},
C\'eline Reyl\'e\inst{2},
and
Mario Ortigoza-Urdaneta\inst{1}
}

\authorrunning{J. G. Fern\'andez-Trincado et al.} 

 \institute{
 	     Instituto de Astronom\'ia y Ciencias Planetarias, Universidad de Atacama, Copayapu 485, Copiap\'o, Chile\\
 	      \email{jfernandezt87@gmail.com and/or jfernandez@obs-besancon.fr}
 	     \and
	   	 Institut Utinam, CNRS UMR 6213, Universit\'e Bourgogne-Franche-Comt\'e, OSU THETA Franche-Comt\'e, Observatoire de Besan\c{c}on, BP 1615, 25010 Besan\c{c}on Cedex, France  
	   	 \and
	   	 Departamento de Astronom\'\i a, Casilla 160-C, Universidad de Concepci\'on, Concepci\'on, Chile
	   	 \and
	   	  Astronomical Institute of the Academy of Sciences of the Czech Republic, Bo\v{c}n\'i II 1401/1, Prague, 141 00, Czech Republic
	   	 \and
	   	  Instituto de Astrof\'{\i}sica de Canarias, V\'{\i}a L\'actea S/N, E-38205 La Laguna, Tenerife, Spain
	   	 \and
          Departamento de Astrof\'{\i}sica, Universidad de La Laguna (ULL), E-38206 La Laguna, Tenerife, Spain
         \and
         School of Physics, University of New South Wales, Sydney NSW 2052
         \and
         Centre of Excellence for Astrophysics in Three Dimensions (ASTRO-3D), Australia
         \and
         Department of Physics and JINA Center for the Evolution of the Elements, University of Notre Dame, Notre Dame, IN 46556, USA
	   	 \and
	   	 Center for Astrophysical Sciences and Department of Physics and Astronomy, The Johns Hopkins University, Baltimore, MD 21218
	   	 \and 
          ELTE E\"otv\"os Lor\'and University, Gothard Astrophysical Observatory, Szombathely, Hungary
	   	  \and
	   	  Premium Postdoctoral Fellow of the Hungarian Academy of Sciences
	   	 \and
	   	 Depto. de Cs. F\'isicas, Facultad de Ciencias Exactas, Universidad Andr\'es Bello, Av. Fern\'andez Concha 700, Las Condes, Santiago, Chile
	   	 \and
	   	 Millennium Institute of Astrophysics, Av. Vicuna Mackenna 4860, 782-0436, Santiago, Chile
	   	 \and
	   	 Vatican Observatory, V00120 Vatican City State, Italy
	   	 \and
	   	 School of Physics and Astronomy, Sun Yat-sen University, Zhuhai 519082, China
	   	 \and
	   	 Universidade de S\~ao Paulo, IAG, Rua do Mat\~ao 1226, Cidade Universit\'aria, S\~ao Paulo 05508-900, Brazil;\\
	   	 }

 \date{Received ...; Accepted ...}
\titlerunning{2MASS J12451043$+$1217401 a N-rich binary in APOGEE-2}

% \abstract{}{}{}{}{} 
% 5 {} token are mandatory

  \abstract
{  We report the serendipitous discovery of a nitrogen-rich, mildly metal-poor ([Fe/H]=-1.08) giant star in a single-lined spectroscopic binary system found in the SDSS-IV Apache Point Observatory Galactic Evolution Experiment (APOGEE-2) survey, Data Release 14 (DR14). Previous work has assumed that the two percent of halo giants with unusual elemental abundances have been evaporated from globular clusters, but other origins for their abundance signatures, including binary mass transfer, must also be explored. We present the results of an abundance re-analysis of the APOGEE-2 high-resolution near-infrared spectrum of 2M12451043+1217401 with the Brussels Automatic Stellar Parameter (BACCHUS) automated spectral analysis code, and re-derive manually the main element families, namely the light elements (C, N), elements (O, Mg, Si), iron-peak element (Fe), \textit{s}-process element (Ce), and the light odd-Z element (Al). Our analysis confirm the N-rich nature of 2M12451043+1217401, which has a [N/Fe] ratio of $+0.69$, and shows that the abundances of C and Al are slightly discrepant from that of a typical mildly metal-poor RGB star, but exhibit Mg, Si, O and \textit{s}-process abundances (Ce) of typical field stars. We also detect a particularly large variability in its radial velocity over the period of the APOGEE-2 observations, and the most likely orbit fit to the radial velocity data has a period of $730.89\pm106.86$ days, a velocity semi-amplitude of $9.92 \pm 0.14$ km s$^{-1}$, and an eccentricity of $\sim 0.1276 \pm0.1174$, which support the hypothesis of a binary companion, and that has probably been polluted by a now-extinct AGB star.
  	}

   \maketitle
  
  %%%%% INTRODUCTION %%%%%%%
  \section{Introduction}
  \label{section1}

       Today it is clear that stellar populations with distinctive light-element abundance patterns \citep{Bastian2018, Fernandez-Trincado2018} are extremely common in globular clusters (GCs), while metal-poor stars ([Fe/H]$\lesssim -0.7$) characterised by enhanced N ([N/Fe]$\gtrsim+0.5$) and depleted C ([C/Fe]$\lesssim+0.15$) are rarely found in the field \citep{Johnson2007, Martell2011, Carollo2013}. While currently still inconclusive, there is tantalising evidence that stars with "anomalous chemistry" may be present beyond GC environments \citep[e.g.,][]{Lind2015, Fernandez-Trincado2016b, Recio-Blanco2017, Fernandez-Trincado2019c}.
       
        To date, there have been a handful of stars fully characterised in terms of their chemistry and the chemical fingerprint of enriched \textit{second population}\footnote{Here, we refer to the second population as the groups of stars showing enhanced Si, N and Al, and depleted C and O abundances, with respect to other field stars at the same metallicity [Fe/H].} stars \citep[e.g.,][]{Martell2016, Fernandez-Trincado2016, Fernandez-Trincado2017L, Schiavon2017, Fernandez-Trincado2019, Fernandez-Trincado2019b}, especially through observations of molecular $^{16}$OH, $^{12}$C$^{14}$N and $^{12}$C$^{16}$O bands in the \textit{H}-band of APOGEE \citep{Majewski2017}, which display the same chemical anomalies as stars in globular clusters, and exhibit conspicuous anomalies of the CNO elements, most notably N. 
        
        These nitrogen-enhanced stars (hereafter N-rich stars) have received significant attention in recent years, primarily because they are believed to be likely relics of surviving Galactic and/or extragalactic \citep[see][for instance]{Fernandez-Trincado2017L} globular clusters \citep{Martell2010}, or now fully dissolved globular clusters \citep{Fernandez-Trincado2015a, Fernandez-Trincado2015b, Reis2018}, and as such, play an important role in deciphering the early history of the Milky Way itself.
        
        A special feature of N-rich stars is their low carbon-abundance ratios, with all stars having [C/Fe] $\lesssim +$0.15 at [Fe/H] $\lesssim -0.7$, and characterised by significant star-to-star variations in the abundances of elements involved in proton-capture reactions, i.e., C, N, O, Mg, and Al. 
    
Empirically, within the population of N-rich stars several subclasses exist, defined by their stellar metallicity and their Al and/or Mg abundances. Most of the N-rich stars in the bulge \citep[e.g.,][]{Schiavon2017} exhibit intermediate aluminum ([Al/Fe]$\lesssim +0.25$) abundance ratios, similar to the thick disk. However, at higher metallicities, [Fe/H]$>-0.7$, \citet{Schiavon2017} identified a second group of N-rich stars chemically distinct from Milky Way stars across a variety of elements. Other groups of N-rich stars have been identified in the inner disk and the halo \citep{Fernandez-Trincado2016, Fernandez-Trincado2017L}, with [Al/Fe] $\gtrsim+0.5$, significantly above the typical Galactic level, across a range of metallicity. These stars are unlikely to have originated in tidally disrupted dwarf galaxies, because of the rarity of Al-rich stars in  current dwarf galaxies \citep{Shetrone2003, Hasselquist2017}. \citet{Fernandez-Trincado2017L} found that some N-rich field stars are also Mg-rich, similar to second-population stars in globular clusters, while some have a factor of two less Mg than field stars at the same metallicity. There is also a subclass of N-rich stars with lower Al abundances, [Al/Fe]$\lesssim+0.1$, which tend to follow halo-like orbits with very little net rotation \citep{Martell2016}. 

The range of elemental abundances that can be derived from APOGEE \textit{H}-band spectra makes it possible to identify and study these classes of stars in detail. With that information we can quantify their occurence rates in the field, and study their overall kinematic properties, in order to better understand their origins.
     
The origins of most of these N-rich stars are currently under investigation, with theories ranging from the formation of a N-rich star via AGB companion mass transfer or a massive evolved massive star, as suggested by \citep{Lennon2003}, where the former AGB companion has since become a faint white dwarf \citep[see][]{Pereira2017}, to early accretion of GCs or dwarf spheroidal galaxies (dSPhs), as such stars appear chemically distinguishable from disk/halo/bulge stars. However, the dynamical history of such stars remains unexplored to date. Measuring additional properties of the N-rich stars may help to further distinguish among them, provide clues to their origins, and/or identify more sub-populations, or more broadly help to understand GC formation and evolution. 
     
     In this study, we present the serendipitous discovery of a N-rich star confirmed to be in a binary system with a compact object. The new object associated to 2M12451043$+$1217401 is a nitrogen-enhanced and carbon-depleted metal-poor star with abundances of Al and Mg mildly discrepant from that of normal RGB stars. We hypothesise that this star is likely to be an example of the result of mass transfer from a binary companion, which is now in the white dwarf stage of stellar evolution. The discovery of such stars helps guide models that attempt to explain the unusual elemental abundances over a wide range of metallicities. 
     
     This paper is outlined as follows. In Section \ref{section2}, we present details and information regarding our serendipitous discovery in the APOGEE-2 dataset. In Section \ref{section4}, we describe and discuss the behaviour of the measured light and heavy elements. In Section \ref{section5}, we present our main conclusions. 
   
%%%%%%%%%%%%%%%%%%%%%%%%%%%%%%%%%%%%%%%%%%%%%%%%%%%%%%
\section{Data}
\label{section2}

For our analysis we make use of the publicly available \textit{H}-band spectra from the APOGEE-2 survey \citep{Majewski2017}. We draw on the latest data release, DR14 \citep{Abolfathi2018}, obtained with the multi-object high-resolution spectrograph APOGEE mounted at the Sloan 2.5m Telescope \citep{Gunn2006} at Apache Point Observatory, which began observation in 2014 as part of the Sloan Digital Sky Survey IV \citep{Blanton2017}. A detailed description of these observations, targeting strategy, data reduction, and APOGEE stellar-parameter estimates can be found in \citet[][]{Zasowski2013}, \citet{Zasowski2017}, \citet{Nidever2015}, \citet{Holtzman2015}, \citet{Zamora2015} and \citet{ASPCAP}, respectively.

We start by searching for high-[N/Fe] outliers in the [N/Fe]-[Fe/H] abundance space in the first Payne data release of APOGEE abundances \citep[see][hereafter \texttt{Payne}-APOGEE]{Payne}. The \texttt{Payne} routine simultaneously derives best-fit values for all atmospheric parameters and abundances using neural networks, with the parameter space of the training set restricted to [Fe/H]$\gtrsim-1.5$. For each source, it only reports values for the measurement with the highest SNR. 

The sample was then restricted to the metallicity regions $-1.5\lesssim$[Fe/H]$\lesssim-0.7$, and giant stars with log $g < 3.6$, which encompasses the transition between the Galactic thick disk and halo, following the same line of investigation as in \citet{Fernandez-Trincado2017L}. Thus, we adopt the atmospheric parameters and [C,N,O/Fe] and [Fe/H] abundance ratios reported in the Payne catalog as the first-guess input parameters into the Brussels Automatic Stellar Parameter (\texttt{BACCHUS}) code \citep{BACCHUS} in order to derive the metallicity and chemical abundances reported in Table \ref{table1}. In addition to those literature values, we also systematically synthesised every element and line of 2M12451043$+$1217401 at high spectral resolution, to compare with the Payne and ASPCAP determinations. We have made careful line selection, as well as providing abundances based on a line-by-line differential approach in the atomic and molecular input data. The abundances that are derived for 2M12451043$+$1217401 using \texttt{BACCHUS}, found in Table \ref{table1}, are generally in good agreement with those in the literature. In Appendix \ref{Append1}, we describe the \texttt{BACCHUS} pipeline in more detail.
  
  \begin{table}[htbp]
  	\begin{tiny}
  		\begin{center}
  			\setlength{\tabcolsep}{1.0mm}  
  			\caption{Comparison of the mean elemental abundances derived from our target 2M12451043$+$1217401 using the “abund” module in \texttt{BACCHUS} code, and those obtained with the Payne and ASPCAP pipeline.}
  			\begin{tabular}{cccc}
  				\hline
  				\hline
  				Element              &   \texttt{BACCHUS}$^{\dagger}$   &  Payne & ASPCAP \\  
  				&     \citet{BACCHUS} & \citet{Payne}  &   \citet{ASPCAP}  \\
  				&       $\sigma_{total} $ &  & \\
  				\hline
  				\hline
  				T$_{\rm eff}$         &                4750.2$\pm$100  K &        4750.2   K  &         4886.9$\pm$92.7 K   \\
  				log $g$                 &                   2.24$\pm$0.3     &          2.24       &            2.22$\pm$0.08     \\
  				$\xi_t$                 &                          1.86$\pm$0.05   km ${\rm s}^{-1}$  &         1.39 km ${\rm s}^{-1}$                &      1.44  km ${\rm s}^{-1}$            \\		  			   
  				${\rm [Fe/H]}$       &          $-$1.08$\pm$0.14    &          $-$1.09      &          $-$1.12$\pm$0.01   \\  	
  				${\rm [C/Fe]}$       &                 0.06$\pm$0.24   &          $-$0.23      &          $-$0.13$\pm$0.07   \\
  				${\rm [N/Fe]}$       &                0.69$\pm$0.22    &               0.66      &            0.74$\pm$0.09   \\
  				${\rm [O/Fe]}$       &                0.46$\pm$0.23    &               0.22      &            0.22$\pm$0.07   \\
  				${\rm [Al/Fe]}$      &                0.04$\pm$0.16    &         $-$0.04      &          $-$0.05$\pm$0.08    \\  	
  				${\rm [Mg/Fe]}$    &                0.10$\pm$0.14     &              0.22       &            0.22$\pm$0.04     \\  	
  				${\rm [Si/Fe]}$       &               0.22$\pm$0.12     &              0.28       &            0.21$\pm$0.04     \\  	 
  				${\rm [Ce/Fe]}$     &               0.49$\pm$0.20     &            ...       &   ... \\
  				\hline
  				\hline
  			\end{tabular}  \label{table1}
  			\tablefoot{The reported uncertainty for each chemical species in column 2 is: $\sigma_{total}  = \sqrt{\sigma^2_{[X/H], T_{eff}}    + \sigma^2_{[X/H],{\rm log} g} + \sigma^2_{[X/H],\xi_t}  + \sigma^2_{mean}   }$. The Solar reference abundances are from \citet{Asplund2005} for light elements and \citet{Grevesse2015} for heavy elements. $^{\dagger}$The \texttt{BACCHUS} pipeline was used to derive the broadening parameters, metallicity, and chemical abundances.}
  		\end{center}
  	\end{tiny}
  \end{table}

\begin{center}
 	\begin{figure*}
 		\includegraphics[width=190mm]{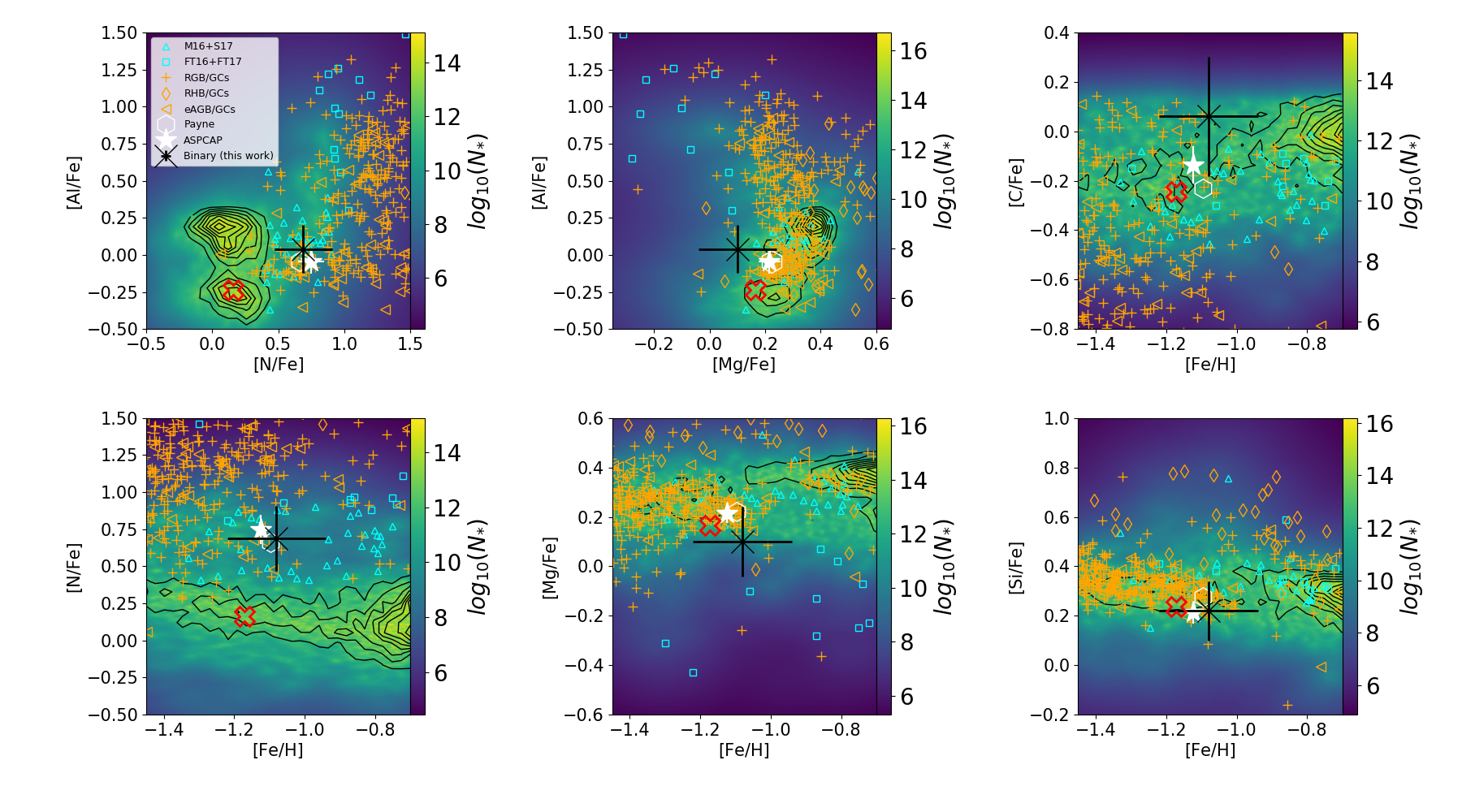}
 		\caption{ The background color in each panel shows a Kernel Density Estimate (KDE) smoothed distribution of [Mg/Fe], [Al/Fe], [N/Fe], [C/Fe], [Si/Fe], and [Fe/H] of Payne-APOGEE stars, while the black contours show the density of stars with "normal" N abundance ($\lesssim+0.5$), and highlighted with a  black symbol is the new N-rich star, 2M12451043$+$1217401 (see text). The plotted error bar shows the classical standard deviation derived from the different abundances of the different lines for each element, as reported in Table \ref{table1}. The same field sample is compared to APOGEE determinations for RGB, eAGB, and RHB stars (orange symbols) in GCs, M13 ([Fe/H]$=-1.53\pm0.04$), M3 ([Fe/H]$=-1.50\pm0.05$), M5 ([Fe/H]$=-1.29\pm0.02$), M107 ([Fe/H]$=-1.02\pm0.02$) and M71 ([Fe/H]$=-0.78\pm0.02$) from \citep[][]{Masseron2018}, and a sample of known N-rich stars plotted using small cyan symbols: unfilled triangles for stars reported in M16--\citet{Martell2016} and S17--\citet{Schiavon2017} from \texttt{Payne}-APOGEE abundances, and unfilled squares for stars studied in FT16--\citet{Fernandez-Trincado2016} and FT17--\citet{Fernandez-Trincado2017L} from a line-by-line differential analysis. The red "X" unfilled symbol mark the typical abundance patterns of a Normal RGB star, 2M12251747$+$1450078, with similar stellar parameters and similar metallicity that 2M12451043$+$1217401.}
 		\label{Figure1}
 	\end{figure*}
 \end{center}

The high-[N/Fe] outlier sample itself was defined by fitting a 5$^{th}$ order polynomial to the run of [N/Fe] with [Fe/H] and taking stars deviating from the fit by $+2.5\sigma$. 2M12451043$+$1217401 is a high-[N/Fe] outlier in chemical space, as compared to the bulk of the disk, bulge, and halo stars in the Payne-APOGEE sample. The sample in Fig. \ref{Figure1} contains stars with [C/Fe]$< +0.15$, because higher [C/Fe] abundance ratios are not typically found in globular clusters, and we want to avoid contamination by CH stars, for instance. 

In this first work, we have only analysed one star (2M12451043$+$1217401), with large variability in radial velocity, out of 35 high-[N/Fe] outliers recently identified in Fern\'andez-Trincado et al., (2019, in prep.), who analysed the same data set. The reader solely interested in the discussion of variability may  see Appendix \ref{Append2} and skip the latter sections. While a comprehensive analysis of our new high-[N/Fe] outliers is beyond the scope of this paper, we did search in the literature for other high-[N/Fe] outliers in the more metal-poor ([Fe/H]$\lesssim-0.7$) population, identified in the APOGEE sample by \citet{Martell2016}, \citet{Schiavon2017}, and \citet{Fernandez-Trincado2016, Fernandez-Trincado2017L}, with available chemical abundances in the \texttt{Payne}-APOGEE catalog, to verify that both Payne-APOGEE abundances and the polynomial fit used in this work is properly returning high-[N/Fe] outliers by recovering the most obvious known nitrogen-enhanced population. The results of this comparison are shown in the left-bottom panel of Fig. \ref{Figure1}. 

\begin{center}
	\begin{figure*}
		\includegraphics[width=195mm]{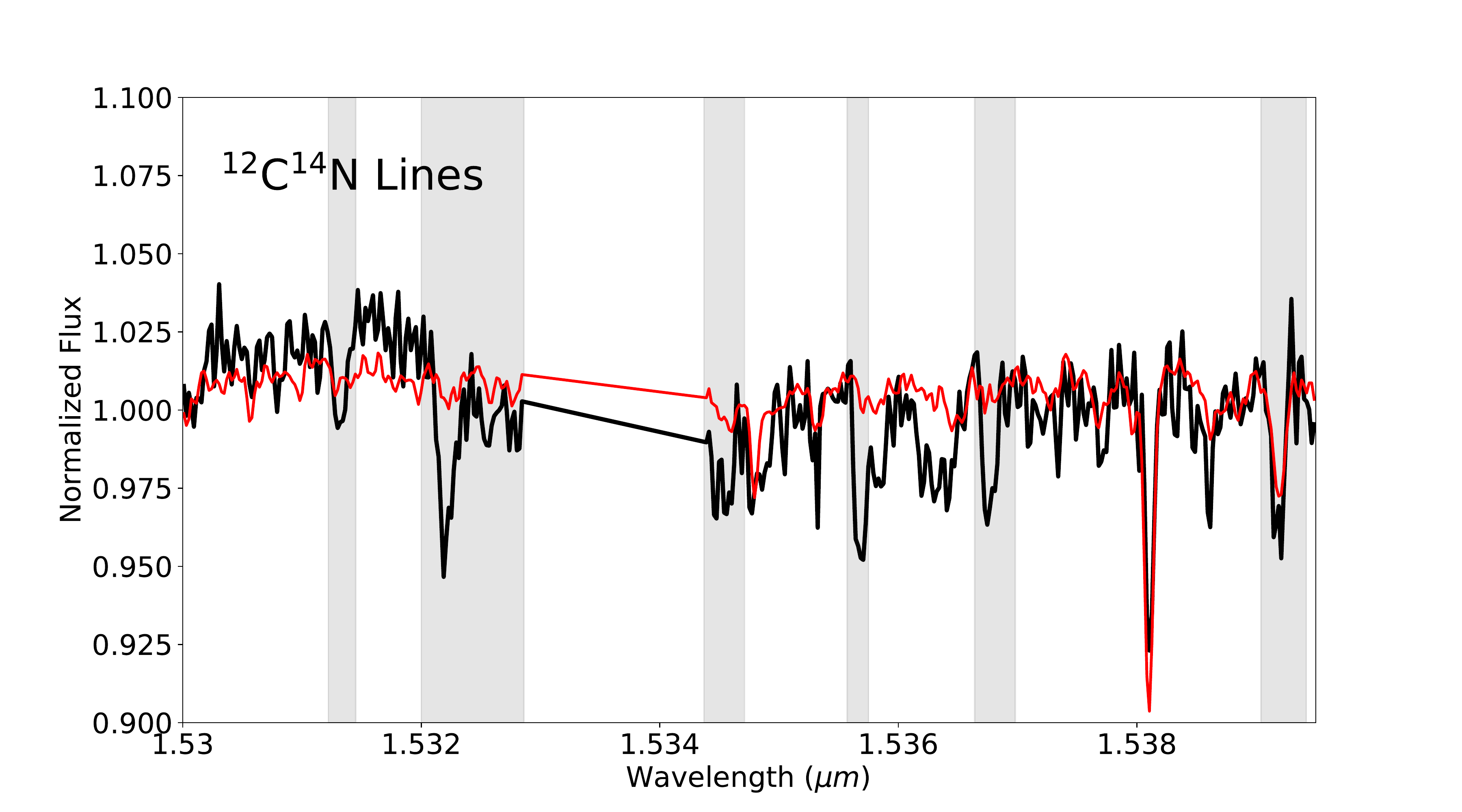}
		\caption{Comparison between the spectrum of a normal (red line, APOGEE\_ID: 2M12251747$+$1450078, T$_{\rm eff} = 4713.9$ K, log \textit{g}$ = $ 2.24, $\xi_t = 1.86$ km s$^{-1}$, [Fe/H]$=-1.17$ dex, and SNR$=637.8$) and our N-rich star (black line), with similar stellar parameters and similar metallicity. The gray vertical bands indicate the positions of $^{12}$C$^{14}$N lines.}
		\label{Figure2}
	\end{figure*}
\end{center}

Figure \ref{Figure2} displays an example for a portion of the observed APOGEE spectrum of 2M12451043$+$1217401, where the $^{12}$C$^{14}$N absorption feature is quite strong. As a comparison, we also show the APOGEE spectrum of a field star with a "normal" nitrogen abundance ([N/Fe]$\lesssim +0.5$), 2M15182930$+$0206378, with stellar parameters and metallicity identical to that of the N-rich star. The N-rich star has remarkably stronger $^{12}$C$^{14}$N lines, which can only mean that it has much higher nitrogen abundance. We conclude that the nitrogen abundance reported here, which is the basis for our identification of a new high-[N/Fe] outlier in the Milky Way, is highly reliable, and detectable in 2M12451043$+$1217401.

 \section{Stellar parameters and chemical-abundance measurements}
\label{section4}

 The first estimates for stellar parameters for 2M12451043$+$1217401 we adopted in the present work are (T$_{\rm eff}$, log \textit{g}, $\xi_t$, [M/H]) = (4750.2 K, 2.24, 1.86 km s$^{-1}$, $-1.09$), taken from \texttt{Payne-}APOGEE catalogue. With this set of parameters, we performed an abundance analysis using the LTE abundance code \texttt{BACCHUS} \citep{BACCHUS}, using the plane-parallel, one dimensional grid of MARCS model atmospheres \citep{Gustafsson2008}. The results are listed in Table \ref{table1}. The metallicity listed in column 2 of Table \ref{table1} is the average abundance of selected Fe I lines, and is in acceptable agreement with that determined from the ASPCAP and Payne-APOGEE pipelines. 
  
 Here, we explore whether or not our chemical-abundance measurements have the ability to chemically tag stars associated with globular clusters by performing a chemical-tagging analysis. Because we are interested in searching for high-[N/Fe] outliers with chemical signatures typical of globular cluster members, we focus on the abundances of C, N, O, Si, Mg, and Al. We have also attempted to measure one additional element, namely Ce II, because it has three promising and detected lines (15784.8, 15958.4 and 16376.5 \AA{}) in the 2M12451043$+$1217401 spectrum (see Figure \ref{Figure9}). This full set of abundance will be the basis of our comparison with the literature values.
 
 \subsection{Light-elements via spectrum synthesis: $^{16}$OH, $^{12}$C$^{16}$O, $^{12}$C$^{14}$N, Mg I, Al I, and Si I lines}
 
 To put our results in context, we combined the chemical-abundance patterns ([Fe/H], [C/Fe], [N/Fe], [Mg/Fe], [Si/Fe], and [Al/Fe]) of 2M12451043$+$1217401, along with the chemical-abundance pattern of globular cluster stars, other unusual giant stars \citep[e.g.,][]{Martell2016, Fernandez-Trincado2016, Fernandez-Trincado2017L, Schiavon2017} from the APOGEE survey, and the entire Milky Way sample, see Figure \ref{Figure1}.
 
 It is immediately clear from Figure \ref{Figure1} that the chemical-abundance pattern of 2M12451043$+$1217401 appears to be distinguishable from Galactic populations. Similarly, its elevated [N/Fe] ratio, as well as the chemical distribution in the $\alpha$-elements, appear most similar to those seen in the chemistry of normal stars in GCs (often called first-generation stars), and comparable to a few nitrogen-enhanced metal-poor field stars at similar metallicity. On the other hand, the moderately enhanced nitrogen level of this star, [N/Fe] $\sim +0.69$, suggests that this value is above the boundary that separates N-rich stars ([N/Fe]$\gtrsim+0.5$) from objects with "normal" N abundances ([N/Fe]$\lesssim +0.5$), according to our strategy to separate these two population (see above text). This is strongly manifested in the [Fe/H] vs. [N/Fe] plane in Figure \ref{Figure1}.

  Figure \ref{Figure1} demonstrates that 2M12451043$+$1217401 seems to resemble the locus dominated by globular cluster stars and the nitrogen-enhanced metal-poor field stars discussed in \citet{Martell2016, Fernandez-Trincado2016, Schiavon2017} and \citet{Fernandez-Trincado2017L} in the  [N/Fe]--[Fe/H] and [Al/Fe]--[Fe/H] planes, which suggest that this star might have originated in globular clusters. However, another possible source for the abundance pattern in 2M12451043$+$1217401, given its radial velocity variability, is binary mass transfer from an possible AGB companion \citep[see, e.g.,][]{Starkenburg2014}. The mass range for the donor star is determined by the minimum mass for the third dredge-up, and by the onset of effective hot bottom burning, which burns C into N quite effectively. { It is important to note that without clear diagnostics such as mass, i.e., from an orbital solution, or a wide gamut of \textit{s}-/\textit{r}-process abundance patterns, the nature of the companion that polluted the N-rich binary star is not at all obvious from observations of [X/Fe] or [X/H] ratios (see below).}
	
 As can be seen in Figure \ref{Figure1}, this star occupies the same region of Mg-Al abundance space as globular cluster stars, without strong evidence of MgAl cycles. The Si abundance is also moderately enriched, which is typical for globular clusters. While the abundance pattern of 2M12451043+1217401 seems consistent with globular cluster stars in the [N/Fe]--[Fe/H], [Mg/Fe]--[Fe/H], [Al/Fe]--[Mg/Fe], and [Al/Fe]--[N/FE], it is distinct from the overall APOGEE data set, which contains bulge, disk and halo stars. We also see some distinction in carbon between 2M12451043+1217401 and the main body of N-normal stars and globular cluster population, with the N-normal stars and GCs stars typically having lower [C/Fe] (for a given metallicity). In Figure \ref{Figure1} we also compared the chemical-composition of a giant star in the main body of N-normal stars with similar stellar parameters and similar metallicity that 2M12451043+1217401, this clearly shows that our object display approximately the same [Ce/Fe] (see below), [Mg/Fe], [Si/Fe], and [Fe/H] abundance ratios that a normal giant star, but show the most significant differences in their [X/Fe] ratios, for $\alpha-$elements, N, C, and Al.

   \begin{center}
   	\begin{figure}
   		\includegraphics[width=95mm]{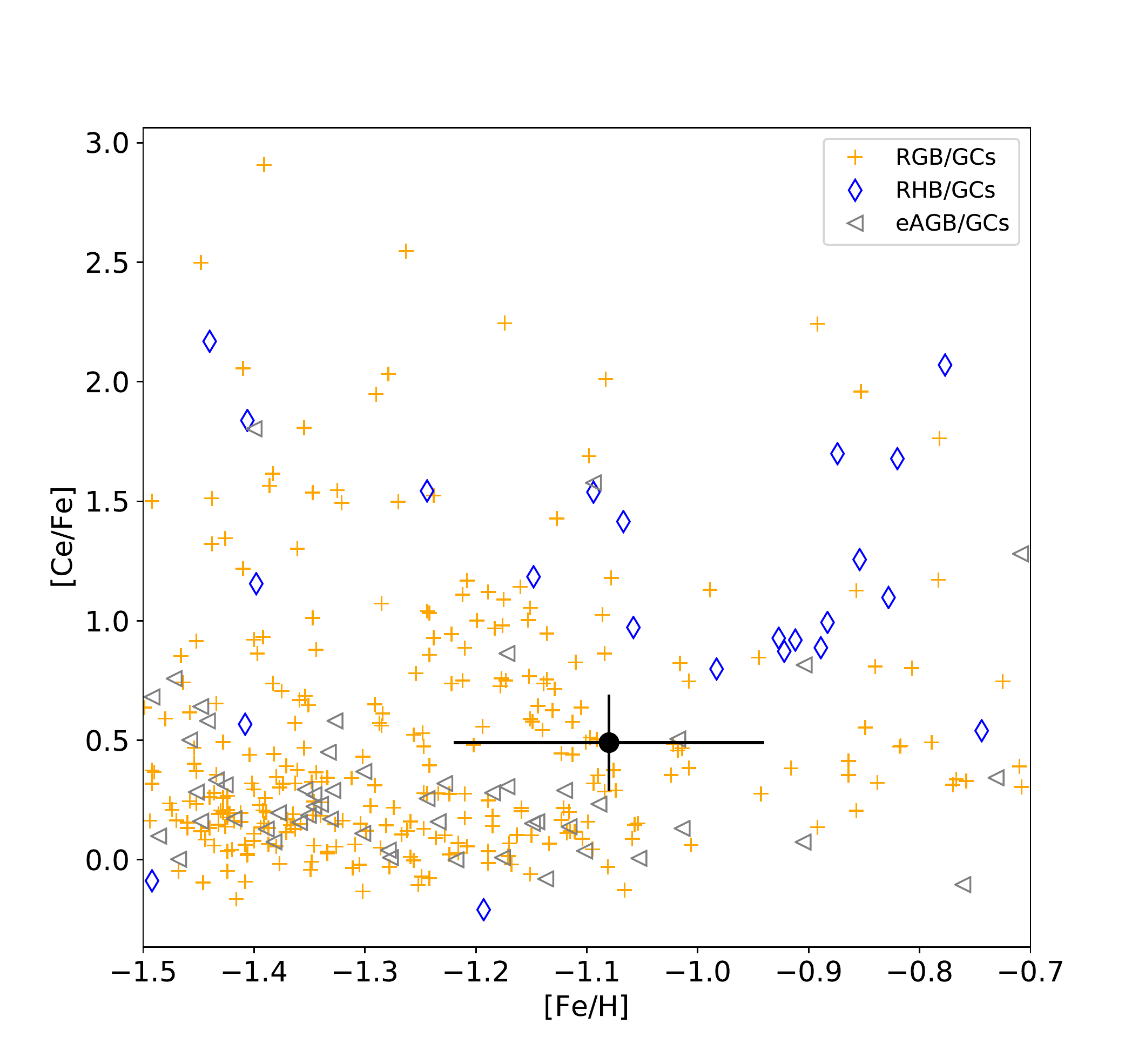}
   		\caption{ Abundances of [Fe/H] vs. [Ce/Fe] in a large number of Galactic globular cluster stars (M107, M71, M5, M3 and M13) from \citet{Masseron2018}. The N-rich binary star (2M12451043$+$1217401) with estimated Ce II indicated by a black filled circle.}
   		\label{FigureCe}
   	\end{figure}
   \end{center}
 
  \subsection{Neutron-capture elements via spectrum synthesis: Ce II lines}
  
\begin{center}
  	\begin{figure*}
  		\includegraphics[width=190mm]{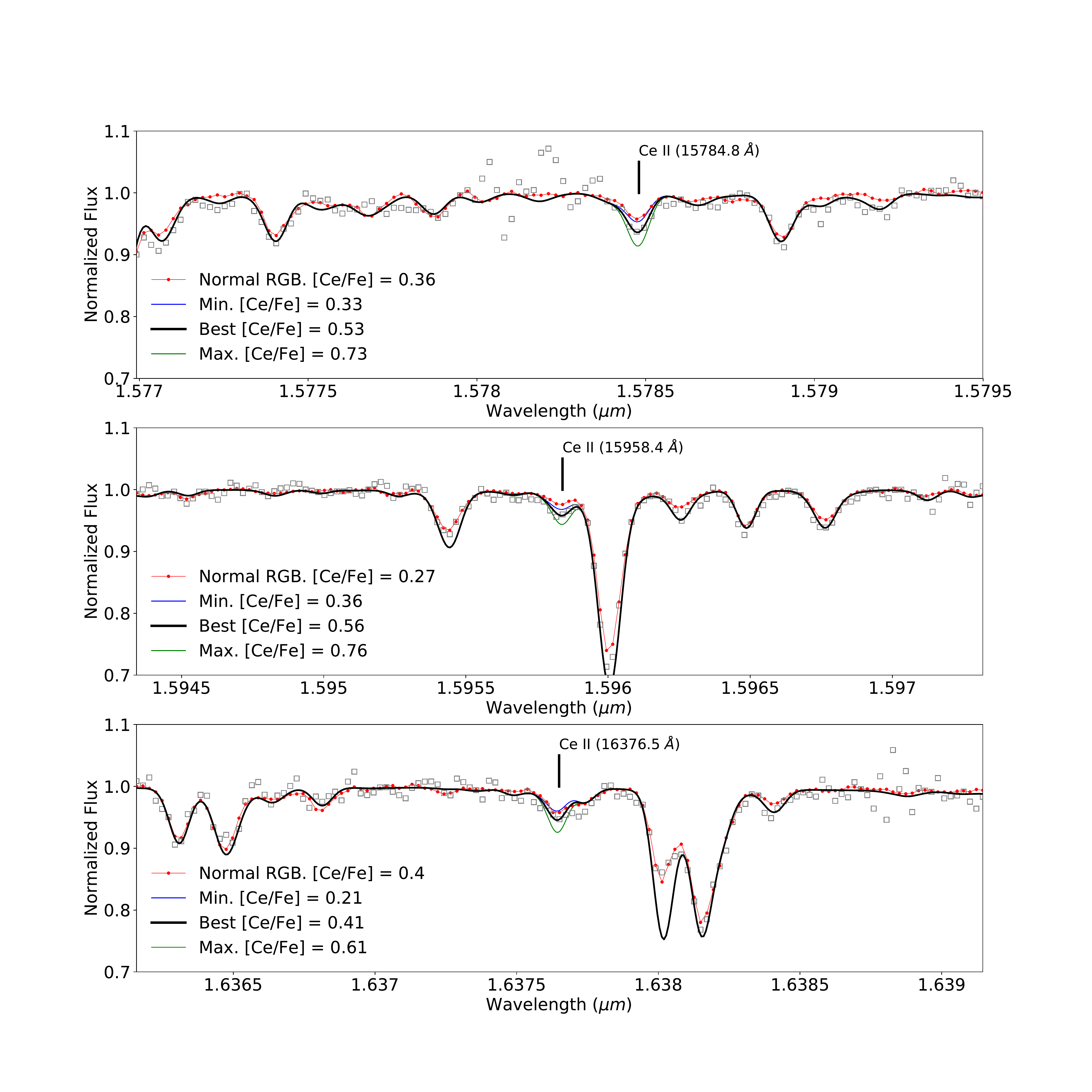}
  		\caption{ Example spectra and the fitted synthesis of Ce II lines for 2M12451043$+$1217401 in the observed infrared spectrum (grey unfilled squares). The printed best fitted abundance (black thick line) values might not be the same as in Table \ref{table1} because the table contains averaged values, not individual fits. The blue and green lines correspond to synthetic spectrum abundance choices that are offset from the best fit by $\pm$0.2 dex. The spectrum of the N-rich star is compared to the spectrum of a normal star with [Ce/Fe]$=+0.34$ dex (red line correspond to APOGEE\_ID: 2M12251747$+$1450078, labeled here as Normal RGB), with similar stellar parameters.}
  		\label{Figure9}
  	\end{figure*}
  \end{center}

\begin{center}
    	\begin{figure}
    		\includegraphics[width=95mm]{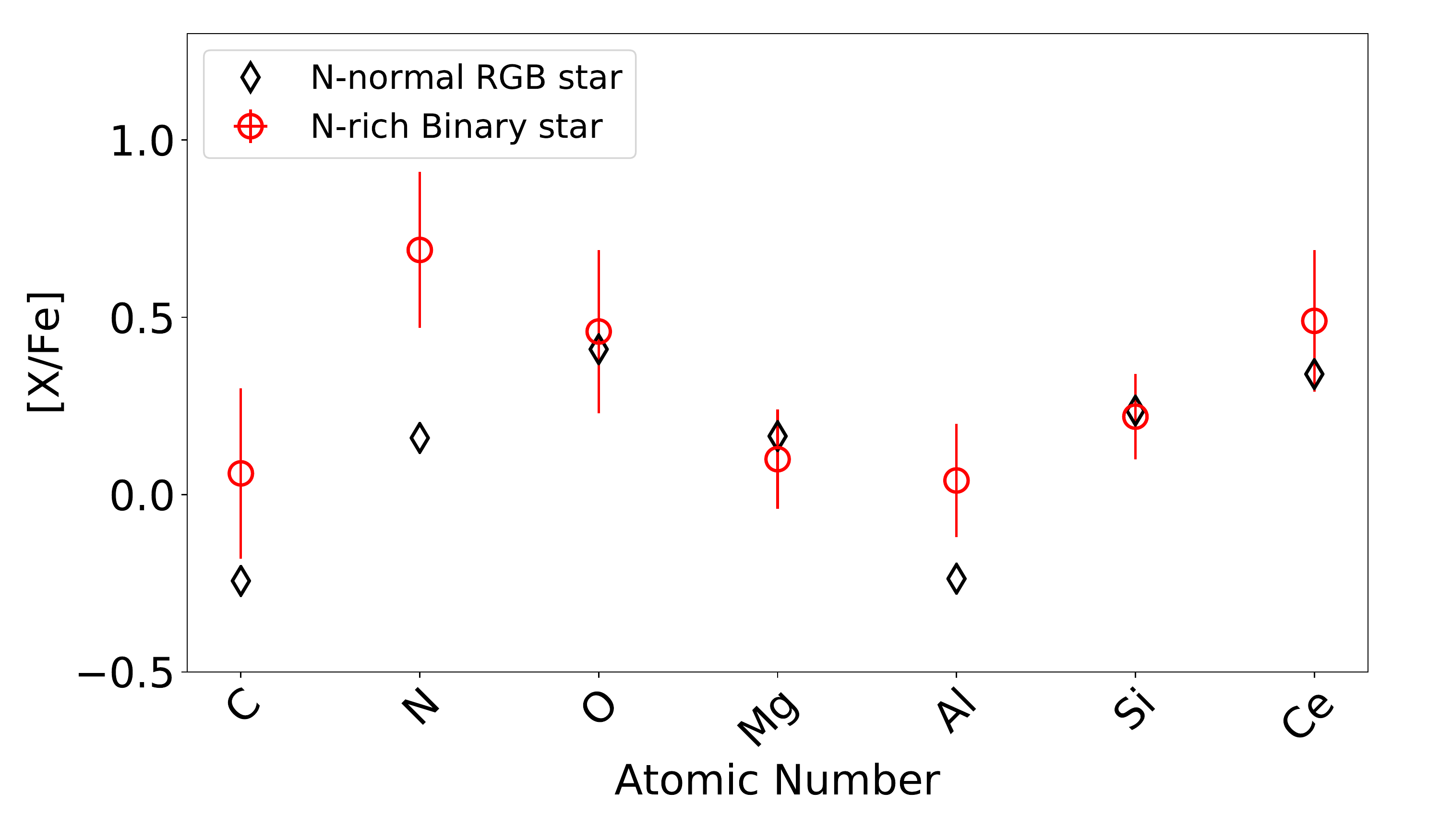}
    		\caption{ The chemical abundance pattern of 2M12451043$+$121740 (N-rich Binary star), for elements X, where X is displayed at the label of the figure. Each determined abundance is shown as an open red circle. These abundances are compared to a N-normal RGB star (black diamond symbols) with similar atmospheric parameters and similar metallicity that the N-rich Binary star.  }
    		\label{Figure8}
    	\end{figure}
    \end{center}

 { Here, from spectral synthesis of the Ce II lines in the APOGEE window, at 15784.8, 15958.4 and 16376.5 \AA{}, we derive a cerium abundance ratio of [Ce/Fe] $= +0.49\pm0.20$, with an uncertainty mainly driven by stellar parameters (see Table \ref{table2}), most sensitive to the surface gravity values. This represents a modest enhancement of the \textit{s}-process in the N-rich binary star, and is similar to the typical enhancement of [Ce/Fe] found in field stars with similar metallicities. 
 	
 	Figure \ref{Figure9} show three of the Ce II lines in 2M12451043+1217401  in the APOGEE window along with the synthetic fits to the spectrum. The Ce II lines are clearly detected and well fit. In the same figure the spectrum of 2M12451043+1217401 is compared to a field (normal) star with similar stellar parameters and with similar metallicity; this brief examination reassures us the existence of a noticeably strengthened in the \textit{s}-process in  2M12451043+1217401, in view of the similarity between to the two stars in all the other relevant parameters, can only mean that it have mildly enhanced values of [Ce/Fe], but still comparable to the \textit{s}-process content of a typical field star (see Figure \ref{Figure8}), and differ only notably on the basis of their nitrogen composition as illustrated in Figures \ref{Figure2} and \ref{Figure8}. 
 	
 	Figure \ref{FigureCe} also show that 2M12451043+1217401 exhibit [Ce/Fe] ratios identical to typical RGB stars as seen in globular cluster environments \citep[e.g.,][]{Masseron2018, Nataf2019} at similar metallicity, with the peculiarity that 2M12451043+1217401 is clearly enhanced in nitrogen as compared to a N-normal field star (see Figure \ref{Figure2}). We showed that 2M12451043+1217401 is a mildly metal-poor binary star characterized by an enhancement of nitrogen with modest enhancement of the \textit{s}-process elements. Therefore, we hypothesize it possible that the high [N/Fe] abundance ratio simultaneous with the basic pattern in Ce, C, and Al, could be due to pollution from a previous AGB companion which is now a white dwarf. 
 
 	} 
  
{ From a theoretical point of view, the intermediate-mass AGB stars, with masses of 3--8 M$_{\odot}$, may influence nucleosynthesis where N is strongly enhanced by Hot-Bottom Burning (HBB) at the expense of C \citep{Masseron2010}, and may be able to produce simultaneously a considerable amount of nitrogen \citep{Cristallo2015}, in this case, the abundances that we measure are not its original ones but they reflect the chemical composition of the companion, plus some degree of dilution with the convective envelope of the accreting star. To summarise, although the mechanism responsible for the N production could be attributed to pollution from an intermediate-mass AGB star, no current AGB models reproduce the trend observed in Figure \ref{Figure8} for 2M12451043$+$1217401. Thus we cannot use N or Ce to place additional constraint on the mass of the progenitor. Thus, a future inventory of the chemistry of this system, in particular, the elements formed by neutron-capture processes, would hint at the range of mass of the companion, and possibly help confirm or refute the association with an extinct AGB star. 

We have shown that 2M12451043$+$1217401 does show radial velocity variation, which is consistent with being a binary. It is indeed possible that an intermediate-mass companion has undergone its AGB phase and dumped shell-nucleosynthesis processed material onto the observed star, in a similar fashion to what happens for CH and CEMP stars \citep[e.g.,][]{Cristallo2016}, increasing the content of N, C, and Al, as compared to a typical field RGB star (see Figure \ref{Figure8}).
}

\section{Concluding remarks}
\label{section5}

In this work, we communicate the serendipitous discovery of an unusual red-giant star that show significantly enhanced [N/Fe] among metal-poor field stars. Based on high-resolution NIR spectroscopic data from the APOGEE-2 survey, we determined the atmospheric parameters, spectrostopic distance, radial velocity variabitity, abundances of light elements (C, N, O, Mg, Al and Si), and the elements created by the \textit{s}-process (Ce II) for 2M12451043$+$1217401. 

Combining the large radial velocity variation, and nitrogen over-abundance, we hypothesise that { an possible AGB-binary system } may produce a [N/Fe] over-abundance in some N-rich stars within the Milky Way, since mass transfer happened in the past, and the star is an normal RGB star where we see the effects due to the pollution of a companion that has undergone its AGB phase and dumped shell-nucleosynthesis processed material onto the observed star, i.e., the AGB companion deposits its N-rich outer layers onto its RGB companion through accreting winds, the binary system produces N (destroying C in the process), that is then mixed throughout the RGB envelope. The companion star in the system is now a possible white dwarf. A future inventory of the chemistry of this binary system, in particular the elements involved in the neutron-capture reactions (i.e., like hs/ls) like that found in the AGB stars, which are independent of the amount of pollution of the observed star undergoes, is crucial to understand the origin of this unique object and place constraint on the range of mass of the companion. 

A dynamical study of 2M12451043$+$1217401 shows that this system has a retrograde motion with a highly eccentric orbit, consistent with the Galactic inner halo population. The results are described in Appendix \ref{Append3} and illustrated in Figure \ref{FigureC1}. By this we mean that this system was formed early on before the inner halo formation, or else, the system formed together with the inner-halo. The present paucity of halo binary N-enriched stars could therefore also have implications for the binary fraction in the field, which to date has been difficult to determine. This study further supports the idea that AGB stars could be the key players in the pollution of $^{12}$C$^{14}$N via a slow stellar wind which has polluted the RGB, thus their atmospheric chemistry reflect some of the yields from an AGB star.

\begin{acknowledgements}  
		We thank the referee for her/his work that improved the manuscript. J.G.F-T is supported by FONDECYT No. 3180210 and the European COST Action CA16117 (ChETEC) project No 41736. R.E.M. acknowledges project fondecyt 1190621, and also acknowledges support by VRID-Enlace 218.016.004-1.0 and the Chilean Centro de Excelencia en Astrof{\'{i}}sica y Tecnolog{\'{i}}as Afines (CATA) BASAL grant AFB-170002. M.C. acknowledges support by Astronomical Institute of the Czech Academy of Sciences through the project RVO 67985815. O.Z. acknowledge support by the MINECO under grant AYA-2017-88254-P. SzM has been supported by the Premium Postdoctoral Research Program of the Hungarian Academy of Sciences, and by the Hungarian NKFI Grants K-119517 and GINOP-2.3.2-15-2016-00003 of the Hungarian National Research, Development and Innovation Office. D.M. acknowledges support from FONDECYT Regular grant No. 1170121, and the kind hospitality of the Osservatorio di Capodimonte/INAF, Italy. T.C.B. and V.M.P. acknowledge partial support for this work from grant PHY 14-30152; Physics Frontier Center / JINA Center for the Evolution of the Elements (JINA-CEE), awarded by the US National Science Foundation. APV acknowledges a FAPESP for the postdoctoral fellowship grant no. 2017/15893-1 and the DGAPA-PAPIIT grant IG100319. B.T. gratefully acknowledges support from National Natural Science Foundation of China under grant No. U1931102 and support from the hundred-talent project of Sun Yat-sen University.
	
	Funding for the \texttt{GravPot16} software has been provided by the Centre national d'\'etudes spatiales (CNES) through grant 0101973 and UTINAM Institute of the Universit\'e de Franche-Comt\'e, supported by the R\'egion de Franche-Comt\'e and Institut des Sciences de l'Univers (INSU). Simulations have been executed on computers from the Utinam Institute of the Universit\'e de Franche-Comt\'e, supported by the R\'egion de Franche-Comt\'e and Institut des Sciences de l'Univers (INSU), and on the supercomputer facilities of the M\'esocentre de calcul de Franche-Comt\'e. 
	
	Funding for the Sloan Digital Sky Survey IV has been provided by the Alfred P. Sloan Foundation, the U.S. Department of Energy Office of Science, and the Participating Institutions. SDSS- IV acknowledges support and resources from the Center for High-Performance Computing at the University of Utah. The SDSS web site is www.sdss.org. SDSS-IV is managed by the Astrophysical Research Consortium for the Participating Institutions of the SDSS Collaboration including the Brazilian Participation Group, the Carnegie Institution for Science, Carnegie Mellon University, the Chilean Participation Group, the French Participation Group, Harvard-Smithsonian Center for Astrophysics, Instituto de Astrof\`{i}sica de Canarias, The Johns Hopkins University, Kavli Institute for the Physics and Mathematics of the Universe (IPMU) / University of Tokyo, Lawrence Berkeley National Laboratory, Leibniz Institut f\"{u}r Astrophysik Potsdam (AIP), Max-Planck-Institut f\"{u}r Astronomie (MPIA Heidelberg), Max-Planck-Institut f\"{u}r Astrophysik (MPA Garching), Max-Planck-Institut f\"{u}r Extraterrestrische Physik (MPE), National Astronomical Observatory of China, New Mexico State University, New York University, University of  Dame, Observat\'{o}rio Nacional / MCTI, The Ohio State University, Pennsylvania State University, Shanghai Astronomical Observatory, United Kingdom Participation Group, Universidad Nacional Aut\'{o}noma de M\'{e}xico, University of Arizona, University of Colorado Boulder, University of Oxford, University of Portsmouth, University of Utah, University of Virginia, University of Washington, University of Wisconsin, Vanderbilt University, and Yale University.
\end{acknowledgements}

%\bibliographystyle{aa}
%\bibliography{references.bib}

\begin{appendix}
	
%%%%%%%%%%%%%%%%%%%%%%%%%%%%%%%%%%%%%%%%%%%%%%%%%%%%%%%%%%%%%%%%

% Supplementary plots and tables

\section{Line-by-line abundances}	
\label{Append1}	

Table \ref{TableA1} and \ref{TableA2} list the atomic and molecular lines used to derived abundances from a line-by-line differential analysis using the the current version of the Brussels Automatic Stellar Parameter (BACCHUS) code \citep[see][]{BACCHUS}, which relies on the radiative transfer code Turbospectrum \citep{Alvarez1998, Plez2012} and the MARCS model atmosphere grid \citep{Gustafsson2008}. For each element and each line, the abundance determination proceeds as in \citet{Hawkins2016}, and summarized here for guidance: (\textit{i}) A spectrum synthesis, using the full set of (atomic and molecular) lines, is used to find the local continuum level via a linear fit; (\textit{ii}) cosmic and telluric rejections are performed; (\textit{iii}) the local S/N is estimated; (\textit{iv}) a series of flux points contributing to a given absorption line is automatically selected; and (\textit{v}) abundances are then derived by comparing the observed spectrum with a set of convolved synthetic spectra characterised by dfferent abundances. Four different abundance determinations are used: (\textit{i}) Line-profile fitting; (\textit{ii}) a core line-intensity comparison; (\textit{iii}) a global goodness-of-fit estimate; and (\textit{iv}) equivalent width comparison. Each diagnostic yields validation flags. Based on these flags, a decision tree then rejects the line or accepts it, keeping the best-fit abundance. We adopted the $\chi^2$ diagnostic for the abundance decision, it is the most robust. However, we store the information from the other diagnostics, including the standard deviation between all four methods. The linelist used in this work is the latest internal DR14 atomic/molecular linelist (linelist.20170418), and the Ce II lines from \citet{Cunha2017}. For a more detailed description of these lines, we refer the reader to a forthcoming paper (Holtzman et al., in preparation). In particular, a mix of heavily CN-cycle and $\alpha$-poor MARCS models were used, as well as the same molecular lines adopted by \citet{Smith2013}, were employed to determine the C, N, and O abundances.  In addition, we have adopted the C, N, and O abundances that satisfy the fitting of all molecular lines consistently; i.e., we first derive $^{16}$O abundances from $^{16}$OH lines, then derive $^{12}$C from $^{12}$C$^{16}$O lines and $^{14}$N from $^{12}$C$^{14}$N, lines and the CNO abundances are derived several times to minimize the OH, CO, and CN dependences \citep[see, e.g.,][]{Smith2013, Fernandez-Trincado2018}.

 We additionally evaluated the possibility to apply the approach of fixing T$^{\rm pho}_{\rm eff}$ and log \textit{g} to values determined independently of spectroscopy, in order to check for any significant deviation in the chemical abundances. For this, the photometric effective temperature, T$^{\rm pho}_{\rm eff} = 4806.6$ K, was calculated from the $J_{\rm 2MASS}-K_{s, 2MASS}$ color relation using the methodology presented in \citet{Gonzalez2009}; for 2M12451043$+$1217401 we adopt $J_{\rm 2MASS}-K_{s, 2MASS}=0.637$ mag and [Fe/H]$=-1.09$. Photometry is extinction-corrected using the Rayleigh Jeans Color Excess (RJCE) method \citep[see][]{Majewski2011}, which leads to $\langle A^{WISE}_{K} \rangle \sim$ 0.023 mag. The resulting temperature, T$^{\rm pho}_{\rm eff} = 4806.6$ K, is in very good agreement with the spectroscopic temperatures from ASPCAP and \texttt{Payne}-APOGEE. In conclusion, this small T$_{\rm eff}$ discrepancy does not affect our results. In the following, our analysis is restricted to the atmospheric parameters as listed in Table \ref{table1}. The same table also lists the abundance measurements in this star obtained with the \texttt{BACCHUS} code and compared to ASPCAP and Payne-APOGEE pipelines. Here, we also list the total error bar on our measurements, which is based on the contributions from the statistical and systematic uncertainties. 

\begin{table}
	\setlength{\tabcolsep}{1.15mm}  
	\begin{tiny}
		\caption{Atomic lines and derived log abundances for the light elements Fe, Al, Mg, and Si, and the heavy element Ce.}
		\begin{tabular}{lcc}
			\hline
			Element  &  $\lambda^{air} ({\rm \AA{}})$ & log ($\epsilon$) \\
			\hline
			{ Fe I} & 15207.5 & 6.194\\
			& 15245.0 & 6.317  \\
			& 15294.6 & 6.225  \\ 
			& 15394.7 & 6.377  \\
			& 15500.8 & 6.210  \\
			& 15501.3 & 6.216  \\
			& 15531.8 & 6.443  \\
			& 15534.2 & 6.415  \\
			& 15588.3 & 6.515  \\ 
			& 15591.5 & 6.308  \\ 
			& 15604.2 & 6.180  \\ 
			& 15621.7 & 6.324  \\
			& 15632.0 & 6.296  \\
			& 15662.0 & 6.332  \\
			& 15723.6 & 6.356  \\ 
			& 15769.1 & 6.583  \\
			& 15769.4 & 6.583  \\
			& 15774.1 & 6.400  \\
			& 15895.2 & 6.456  \\
			& 15904.3 & 6.419  \\
			& 15920.6 & 6.324  \\
			& 15964.9 & 6.460  \\
			& 15967.7 & 6.414  \\
			& 15980.7 & 6.331  \\
			& 16006.8 & 6.356  \\
			& 16007.1 & 6.357  \\
			& 16040.7 & 6.437  \\
			& 16042.7 & 6.172  \\
			& 16071.4 & 6.453  \\
			& 16125.9 & 6.505  \\
			& 16153.2 & 6.419  \\
			& 16179.6 & 6.379  \\
			& 16195.1 & 5.969  \\
			& 16284.8 & 6.489  \\
			& 16517.2 & 6.402  \\
			& 16524.5 & 6.427  \\
			& 16561.8 & 6.483  \\
			& 16645.9 & 6.300  \\
			& 16665.5 & 6.468  \\
			${\rm \langle A(Fe) \rangle }\pm \sigma_{mean}$ & & 6.37$\pm$0.12\\
			\hline
			{ Al I} 	& 16719.0 & 5.253 \\
			& 16750.6 & 5.237 \\
			& 16763.4 & 5.503 \\
			${\rm \langle A(Al) \rangle }\pm \sigma_{mean}$ & & 5.33$\pm$0.12 \\
			\hline
			{ Mg I} 	& 15740.7 & 6.586 \\
			& 15748.9 & 6.556 \\
			& 15765.8 & 6.537 \\
			${\rm \langle A(Mg) \rangle }\pm \sigma_{mean}$ & & 6.56$\pm$0.02\\
			\hline
			{ Si I}  & 15376.8 & 6.450 \\
			& 15557.8 & 6.629 \\
			& 15884.5 & 6.475 \\
			& 15960.1 & 6.623 \\
			& 16060.0 & 6.738 \\
			& 16094.8 & 6.728 \\
			& 16215.7 & 6.762 \\
			& 16241.8 & 6.775 \\
			& 16680.8 & 6.570 \\
			& 16828.2 & 6.766 \\			 
			${\rm \langle A(Si) \rangle }\pm \sigma_{mean}$ &  & 6.65$\pm$0.12   \\
			\hline
			{ Ce II} & 15784.8   &   1.029 \\
			& 15958.4 &  1.055 \\
			& 16376.5 & 0.911 \\
			${\rm \langle A(Ce) \rangle }\pm \sigma_{mean}$ & &0.998$\pm$0.06 \\
			\hline
		\end{tabular}  \label{TableA1}
	\end{tiny}
\end{table}

\begin{table}
	\setlength{\tabcolsep}{1.15mm}  
	\begin{tiny}
		\caption{Molecular features and log abundances used to derive C, N, and O.}
		\begin{tabular}{lcc}
			\hline
			Element  &  $\lambda^{air} ({\rm \AA{}})$ & log ($\epsilon$)  \\
			\hline
			{ $^{12}$C from $^{12}$C$^{16}$O lines} 	& 15578.0 & 7.532 \\
			& 16185.5 & 7.202 \\
			${\rm \langle A(C) \rangle }\pm \sigma_{mean}$       &                               &  7.37$\pm$0.17 \\
			\hline
			{ $^{14}$N from $^{12}$C$^{14}$N lines}	& 15158.0 & 6.980 \\
			&			15165.4 & 7.483 \\
			&			15185.6 & 7.152 \\
			&			15210.2 & 7.228 \\
			&			15222.0 & 7.433 \\
			&			15251.8 & 7.118 \\
			&			15284.5 & 7.494 \\
			&			15309.0 & 7.442 \\
			&			15317.6 & 7.591 \\
			&			15328.4 & 7.226 \\
			&			15363.5 & 7.483 \\
			&			15447.0 & 7.432 \\
			&			15462.4 & 7.400 \\
			&			15466.2 & 7.587 \\
			&			15495.0 & 7.489 \\ 
			&			15514.0 & 7.420 \\
			&			15581.0 & 7.264 \\
			&			15659.0 & 7.592 \\
			&			16244.0 & 7.507 \\
			${\rm \langle A(N) \rangle }\pm \sigma_{mean}$       &           &   7.39$\pm$0.17   \\
			\hline
			{ $^{16}$O from $^{16}$OH lines}  & 15409.2 & 8.153 \\
			&	15568.8 & 8.003 \\
			&	16052.8 & 7.748 \\
			&	16055.5 & 7.660 \\
			&	16255.0 & 7.935 \\
			&	16312.6 & 8.287 \\
			&	16534.6 & 7.893 \\
			&	16650.0 & 8.149 \\
			&	16656.0 & 8.160 \\
			&	16704.5 & 8.215 \\
			&	16898.9 & 8.153 \\
			${\rm \langle A(O) \rangle }\pm \sigma_{mean}$ &                    & 8.03$\pm$0.19 \\
			\hline
		\end{tabular}  \label{TableA2}
	\end{tiny}
\end{table}

\begin{table}[htbp]
	\begin{center}
		\setlength{\tabcolsep}{2.0mm}  
		%  	\begin{tiny}
		\caption{Abundance determination sensitivity to the stellar parameters from our present measurements.}
		\begin{tabular}{cccc}
			\hline
			\hline
			$\Delta$[X/H]       &   T$_{\rm eff}\pm100$ K   &  log $g\pm0.3$ & $\xi_t\pm0.05$ km s$^{-1}$ \\  
			\hline
			\hline
			Fe          &   0.03  & 0.05  & 0.002  \\
			C           &   0.14  & 0.06  &  0.002 \\   
			N           &   0.11  & 0.09  & 0.001  \\           
			O           &   0.09 & 0.03  &  0.007 \\ 
			Al           &  0.04  & 0.04 &  0.004 \\                     
			Mg         &   0.08 & 0.09  &  0.060 \\    
			Si           &   0.02 & 0.06  & 0.004  \\   
			% 			Ti           &  0.12  & 0.04  &  0.002 \\
			Ce         &   0.05  & 0.19  & 0.002  \\               
			\hline
			\hline
		\end{tabular}  \label{table2}
		%\tablefoot{}
		%  	\end{tiny}
	\end{center}
\end{table}

%%%%%%%%%%%%%%%%%%%%%%%%%%%%%%%%%%%%%%%%%%%%%%%%%%%%%%%%%%%%%%%%
% Section Variability 
%%%%%%%%%%%%%%%%%%%%%%%%%%%%%%%%%%%%%%%%%%%%%%%%%%%%%%%%%%%%%%%%

\section{Variability} 
\label{Append2}	

\begin{table}
	\setlength{\tabcolsep}{2.0mm}  
	\caption{APOGEE Observations}
	\label{Variability}
	\begin{tabular}{cccc}
		\hline
		\hline
		Julian date  & RV (km s$^{-1}$) & $\sigma$(km s$^{-1}$) & SNR (pixel$^{-1}$)\\
		\hline
		\hline
		2457059.94391  & $-$88.85 & 0.16  & 20  \\
		2457060.91321  & $-$89.43 & 0.15  & 20  \\
		2457062.99631  & $-$89.45 & 0.15  & 20  \\
		2457064.89865  & $-$90.54 & 0.17  & 18  \\
		2457114.79574  & $-$92.80 & 0.22  & 17  \\
		2457118.85652  & $-$92.28 & 0.43  &  8  \\
		2457121.82661  & $-$92.55 & 0.19  & 16  \\
		2457122.80708  & $-$93.23 & 0.14  & 23  \\
		2457141.73865  & $-$93.40 & 0.21  & 15  \\
		2457142.73487  & $-$93.09 & 0.17  & 18  \\
		2457148.68704  & $-$94.02 & 0.19  & 16  \\
		2457167.65509  & $-$92.67 & 0.79  &  5  \\
		2457449.88122  & $-$74.01 & 0.12  & 26  \\
		2457465.78237  & $-$74.55 & 0.12  & 22  \\
		2457468.79933  & $-$73.70 & 0.14  & 22  \\
		2457472.79234  & $-$74.23 & 0.14  & 23  \\
		2457473.88693  & $-$73.25 & 0.16  & 20  \\
		2457475.83189  & $-$73.32 & 0.13  & 22  \\
		2457492.72504  & $-$73.91 & 0.12  & 23  \\
		2457496.80519  & $-$73.52 & 0.21  & 14  \\
		2457499.79405  & $-$73.47 & 0.16  & 19  \\
		\hline
		2457504.78284  & $-$69.84 & 0.89  &  4  \\
		\hline
		2457530.70246  & $-$73.99 & 0.19  & 16  \\
		2457533.71368  & $-$73.81 & 0.11  & 25  \\
		2457534.73149  & $-$74.29 & 0.19  & 14  \\
		\hline
		\hline
	\end{tabular} 
	\tablefoot{$^{\dagger}$Reduced Heliocentric JD. $^{\ddagger}$Values omitted for spectra with SNR < 5.}
\end{table}

Here we use multi-epoch radial velocity measurements available in the APOGEE-2 DR14 database. 2M12451043$+$1217401 was observed multiple times in a series of 25 "visits" in order to meet the signal-to-noise ratio requirements of the APOGEE-2 survey. The radial velocities for each visit are determined using an iterative scheme, i.e., the individual visit spectra are combined using initial guesses for the relative radial velocities into a co-added spectrum, which is then used to re-derive the relative visit velocities \citep[see, e.g.,][]{Nidever2015, PriceWhelan2018}. 

Generally, observations from the APOGEE-2 survey have a relatively short ($\lesssim$ 6 months) baseline, which is a potential limitation to enable the detection of N-rich stars formed through the binary channel. Here, however, we find that the large scatter ($>9.92$ km s$^{-1}$) shown by the measured radial velocities in 2M12451043$+$1217401 implies the clear existence of at least one companion. So far, no evidence of strong radial velocity variations has been found in stars with nitrogen over-abundances in the APOGEE survey \citep[see, e.g.,][]{Martell2016, Fernandez-Trincado2016, Fernandez-Trincado2017L, Schiavon2017}. Therefore, establishing the presence of radial velocity variation among these stars would be required in order to understand if many, or all such objects, formed through the binary channel. 

Although APOGEE-2 observed our star during multiple "visits", we omitted one observation with a signal to noise ratio (SNR) below 4 per pixel from our radial velocity analysis. The observations are summarized in Table \ref{Variability}, which lists the APOGEE radial velocities, their uncertainties, and the signal-to-noise ratio (SNR) per pixel for each epoch. 

\begin{center}
	\begin{figure*}
		\includegraphics[width=170mm]{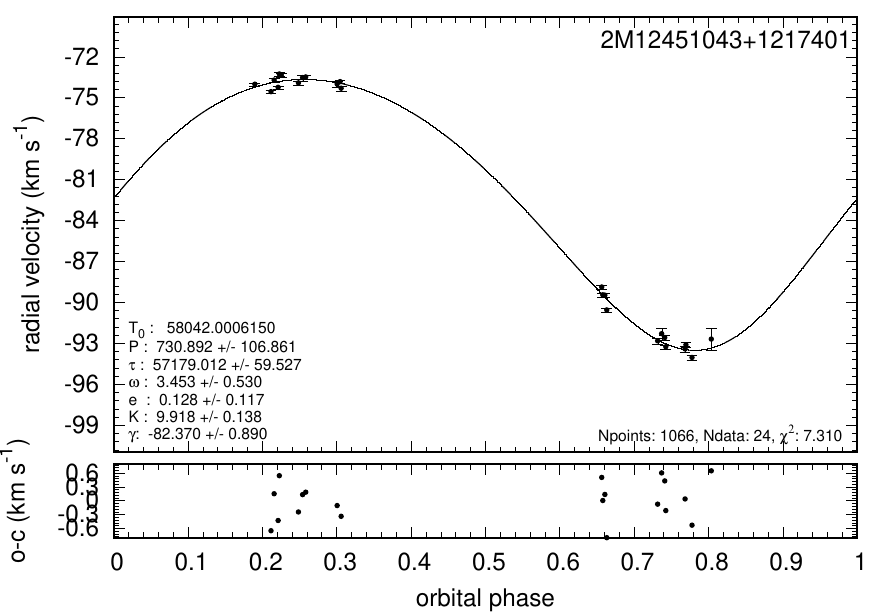}
		\caption{Radial velocity measurements for 2M12451043$+$1217401 from 24 visits of the APOGEE-2 survey (\textit{top panel}) with the best-fit, with the residual velocities (\textit{bottom panel}).}
		\label{figure2}
	\end{figure*}
\end{center}

We conducted a search for the orbital period using the genetic algorithm \texttt{PIKAIA} \citep{Charbonneau1995} to determine the orbital parameters that best fit the available data. Following the analysis described by \citet{Mennickent2012, Mennickent2018, Mennickent2018b} we minimize $\chi^2$ defined as:\\

\begin{equation}\label{chi2}
\chi^2(P,\tau,\omega,e,K,\gamma) =\frac{1}{N-6}\sum_{j=1}^{n}\left(\frac{V_J-V(t_j,P,\tau,\omega,e,K,\gamma)}{\sigma_j}\right)^2,
\end{equation}

\noindent
where $N$ is the number of observations, and $V_j$ and $V$ are the observed and calculated velocities in time $t_j$. The fit velocity is:\\

\begin{equation}\label{eqn:vt}
V(t)=\gamma + K ((\omega+\theta(t)) + e\cos(\omega)),
\end{equation}

\noindent
where $\theta$ is the true anomaly obtained by solving the equations:\\

\begin{equation}
\tan\left(\frac{\theta}{2}\right) = \sqrt{\frac{1+e}{1-e}}\tan\left(\frac{E}{2}\right),
\end{equation}

\begin{equation}
E - e \sin(E)= \frac{2\pi}{P}(t-\tau),
\end{equation}

\noindent
where $E$ is the eccentric anomaly. We constrained the eccentricity between 0 and 1, $\omega$ between $0$ and $2\pi$,
$\tau$ between the minimum HJD and this value plus the period, 
$K$ between $0$ and $(V_{max} - V_{min})$, and $\gamma$ between $V_{min}$ and $V_{max}$.

In order to estimate the errors for the results obtained from \texttt{PIKAIA}, we proceeded to calculate the confidence intervals for the region corresponding to 68.26\% of the sample (1$\sigma$). The results are listed in Table \ref{tab:pikresult}. The fit produces a good match to the available data (residuals $\approx$ 1 km s$^{-1}$, see bottom panel in Fig. \ref{figure2}). 

\begin{table}
	\centering
	\setlength{\tabcolsep}{3.0mm}  
	\begin{tiny}  	
		\caption{Orbital elements for the donor of 2M12451043$+$1217401 obtained by minimization of the $\chi^2$ parameter given by Eq. \ref{chi2}. The value $\tau^*=\tau-2450000$ and the limits of the confidence intervals within one standard deviation ($1\sigma$) are given.}
		\label{tab:pikresult}
		\begin{tabular}{cccc}
			\hline
			Parameter&Best value&Lower limit&Upper limit\\
			\hline
			$P_o$ (d) &730.89 $\pm$ 106.86 & 679.61 &893.33\\
			$\tau^{*}$ &57179.0117 $\pm$  59.5272 &57129.9301 &57248.9845\\
			$\omega$ & 3.45 $\pm$   0.53 [rad] & 2.961 &4.022 \\
			$e$ &0.1276 $\pm$ 0.1174 &0.0569 &0.2916\\
			$K_2$ (km s$^{-1}$) & 9.92 $\pm$  0.14 & 9.856 &10.131 \\
			$\gamma$ (km s$^{-1}$)& -82.37 $\pm$ 0.89 &-83.22 &-81.44\\
			\hline
		\end{tabular}
	\end{tiny}
\end{table}

It is important to note that 2M12451043$+$1217401 was identified as a "bimodal binary" in the bimodal period samplings of \textit{The Joker} code in \citet[][]{PriceWhelan2018}, correponding to periods between $P\simeq$ 650--950 days and eccentricity between $\simeq$ 0.07--0.3: ($P_1, e_1 $) = ($939.80$, $0.31$) days and ($P_2, e_2 $) = ( $689.85$ days, $0.07$). The best-fit orbital parameters from \texttt{PIKAIA} are an orbital period of 730.89 $\pm$ 106.86 days, a velocity semi-amplitude of $9.92 \pm 0.14$ km s$^-1$, and an eccentricity of $0.1276 \pm 0.1174$. These are in acceptable agreement with the shorter period reported by \citet[][]{PriceWhelan2018}.

Visual inspection of the spectrum of 2M12451043+1217401 does not reveal any obvious signs of binary interaction, such as emission lines from an accretion disk. If this is a post-mass-transfer binary system, the primary would have evolved into a white dwarf by this point, making it quite difficult to detect in the \textit{H}-band. Alternately, if there was very strong mass loss from the primary, the binary system could have disrupted or significantly widened. 
For the case of SB1 binaries, where only one component is detected in the spectrum, the mass information is contained in a single function, called the mass function, defined as:

\begin{equation}\label{functionmasss}
f = \frac{(m_1^3)(sin^3(i))}{(m_1 + m_2)^2} = 1.0361 \times 10^{-7} (1-e^2)^{3/2}   \left(\frac{K_2}{km s^{-1}} \right)^3 \frac{P_o}{days} M_{\odot}
\end{equation}

\begin{center}
	\begin{figure}
		\includegraphics[width=90mm]{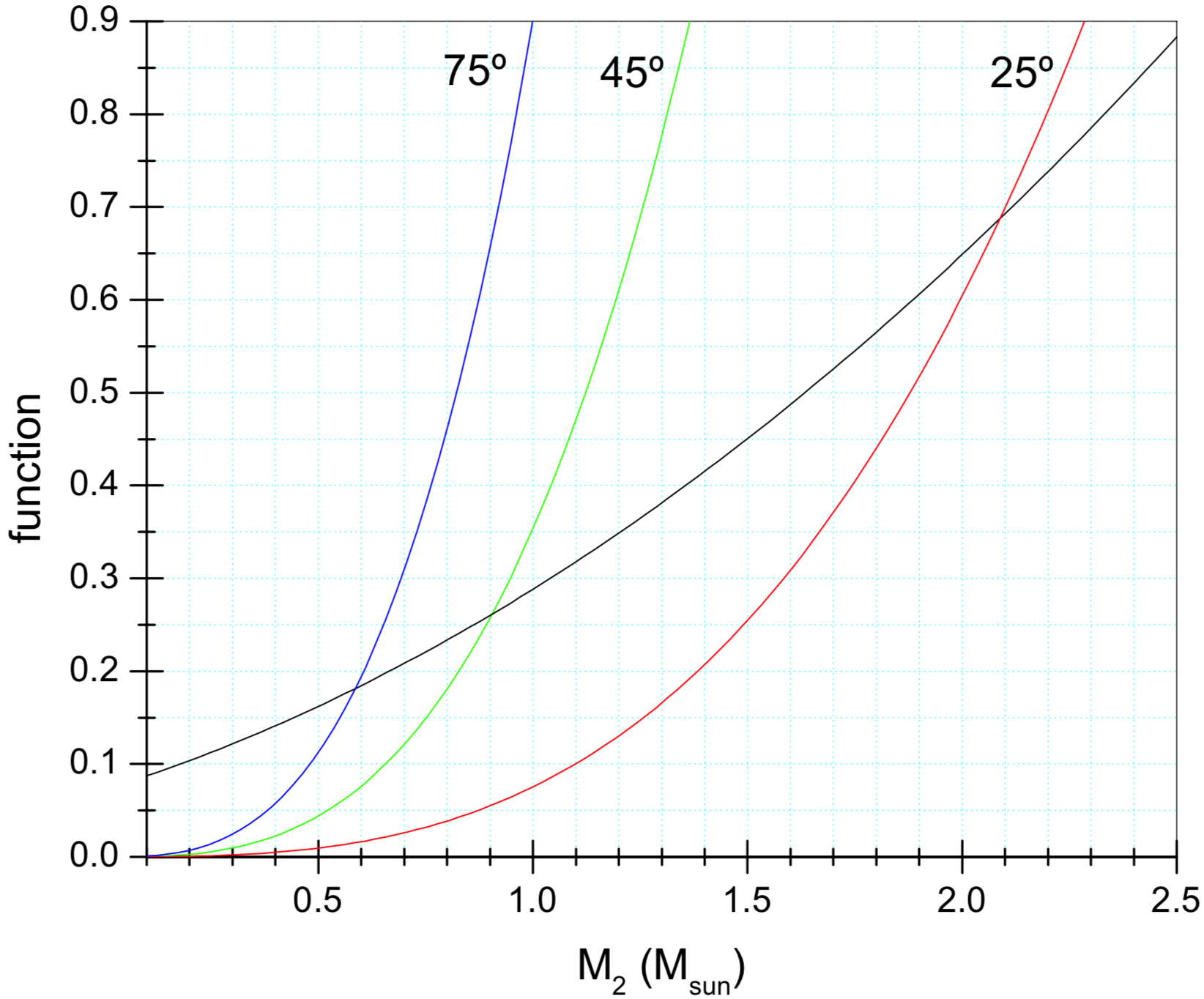}
		\caption{Estimated mass of the unseen companion as a function of $\sin i$.}
		\label{Figuremass}
	\end{figure}
\end{center}

This corresponds to the minimum mass of the unseen companion, labeled with subscript 1 here, in this case 0.07 M$_{\odot}$. Since the system inclination is unknown, several solutions are possible for the mass of the unseen companion. Assuming 1 M$_{\odot}$ for the detected star \citep[][]{PriceWhelan2018}, we summarize our restrictions for the mass of the unseen companion in Fig. \ref{Figuremass}.

From the system mass function (Eq. \ref{functionmasss}), and assuming a 1 M$_{\odot}$ for the detected component, we can determine the mass of the unseen companion from the intersection of two functions of the form $m_1^{3} \sin^3i$ and $g(P, K_2, e) \times (m_1 + m_2)^{2}$, where $g$ is a function of the orbital period, radial velocity half-amplitude, and orbital eccentricity. These functions are plotted for our system for three different angles showing possible range of masses for the companion (see Figure \ref{Figuremass}).

\section{Dynamical behaviour}	
\label{Append3}	

Here we present a first attempt to predict the probable orbit of the newly discovered N-rich binary across the Milky Way. To do this, we have used a state-of-the art orbital integration model in an (as far as possible) realistic gravitational potential, that fits the structural and dynamical parameters of the Galaxy based on the recent knowledge of our Milky Way. For the computations in this work, we have employed the rotating "boxy/peanut" bar model of the novel galactic potential model, called \texttt{GravPot16}\footnote{\url{https://fernandez-trincado.github.io/GravPot16/}}, along with other composite stellar components. The considered structural parameters of our bar model, e.g., mass, present-day orientation, and pattern speeds, is within observational estimations that lie in the range of 1.1$\times$10$^{10}$ M$_{\odot}$, 20$^{\circ}$, and 35--50 km s$^{-1}$ kpc, respectively. 

\begin{center}
	\begin{figure*}
		\includegraphics[width=95mm]{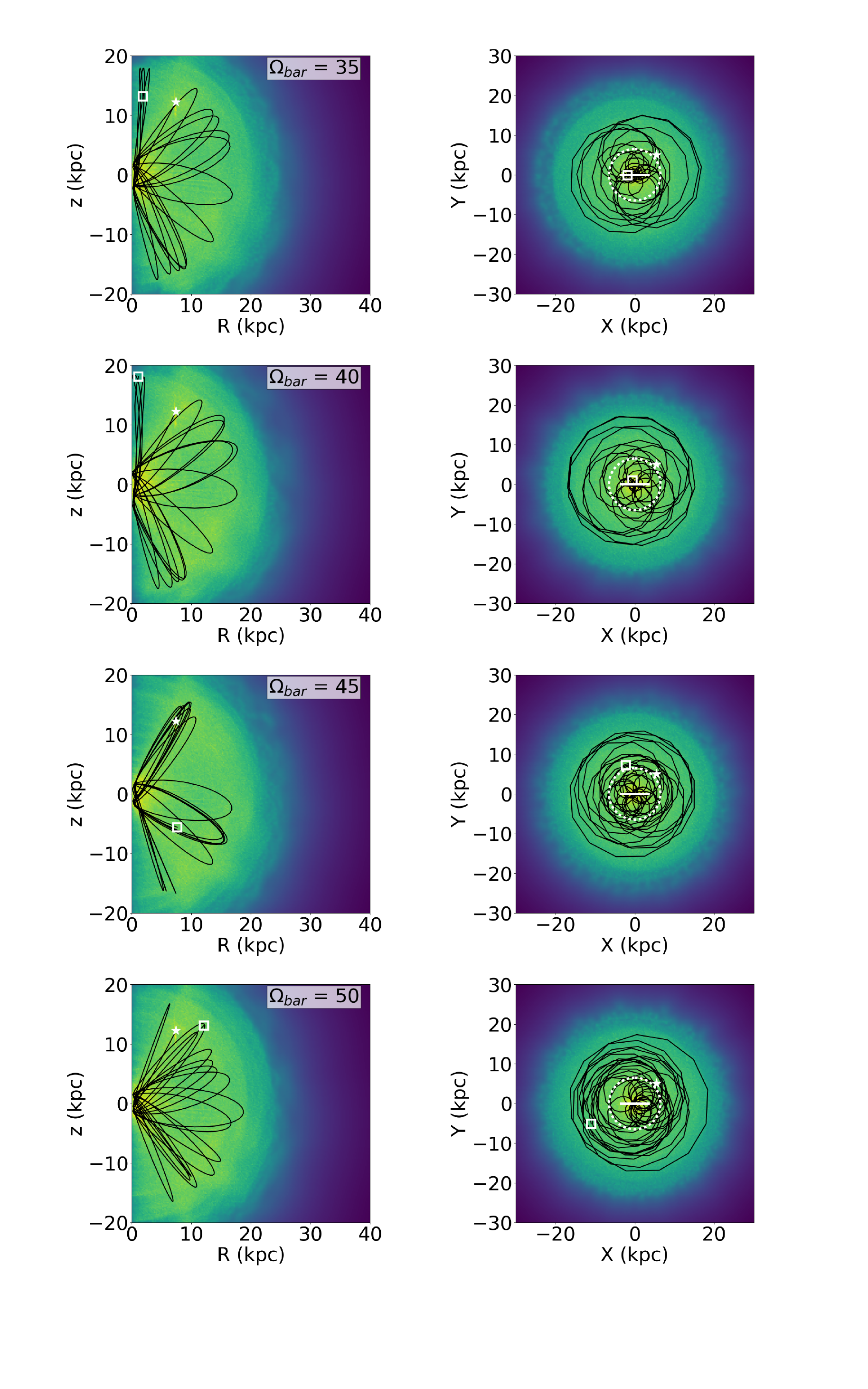}\includegraphics[width=95mm]{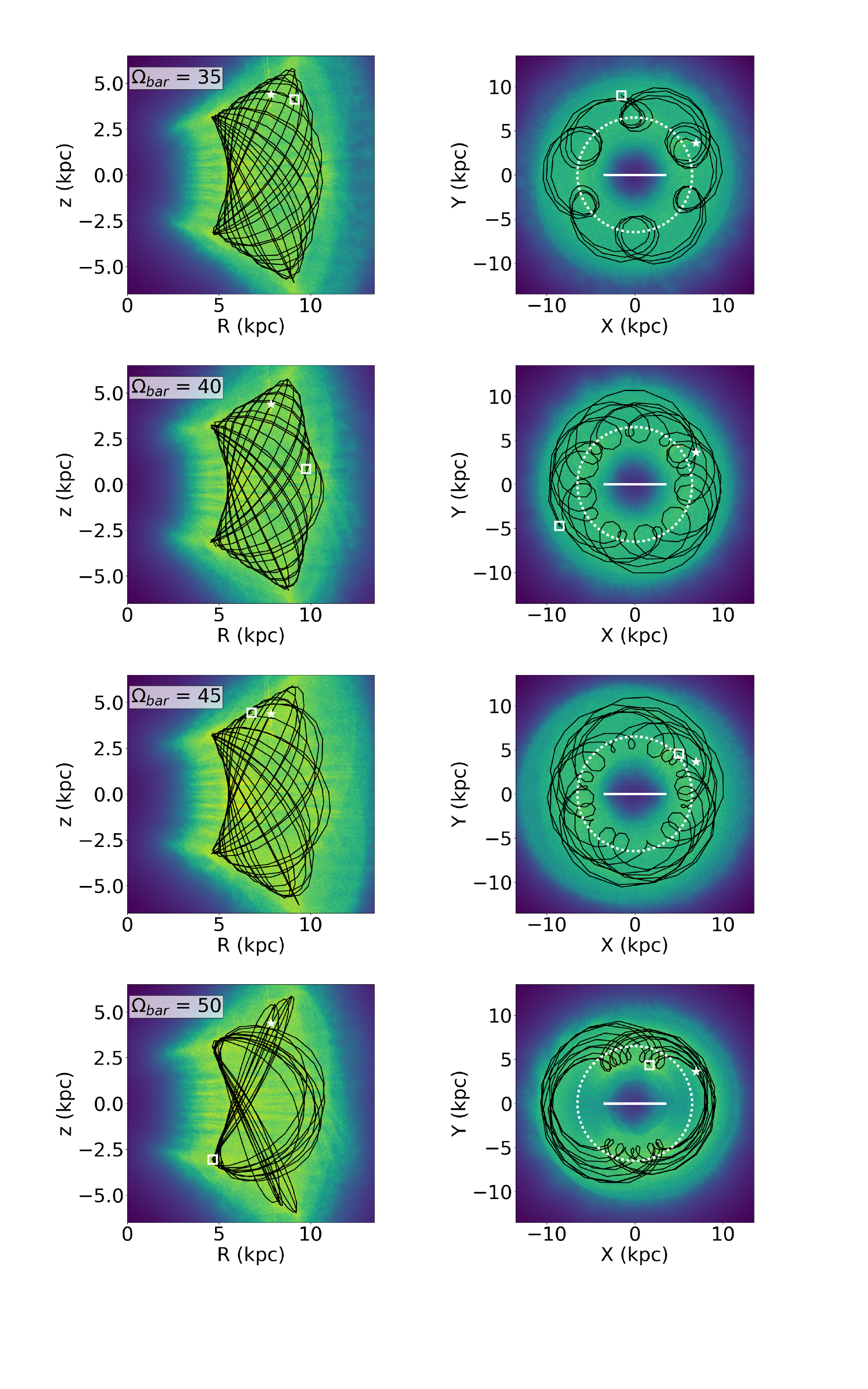}
		\caption{Kernel Density Estimate (KDE) smoothed distribution of simulated orbits employing a Monte Carlo approach, showing the probability densities of the resulting orbits projected on the equatorial and meridional Galactic planes in the non-inertial reference frame where the bar is at rest. The green and yellow colors correspond to more probable regions of the space, which are crossed more frequently by the simulated orbits. The black line is the orbit of 2M12451043$+$1217401 adopting the central inputs. The small white star marks the present position of the cluster, whereas the white square marks its initial position. In all orbit panels, the white dotted circle show the location of the co-rotation radius (CR), the horizontal white solid line shows the extension of the bar. Columns 1 and 2 show the results adopting the estimated distance from the \texttt{StarHorse} code, while columns 3 and 4 for estimated distance from \citet{jones18}.}
		\label{FigureC1}
	\end{figure*}
\end{center}

For reference, the Galactic convention adopted by this work is: $X-$axis is oriented toward $l=$ 0$^{\circ}$ and $b=$ 0$^{\circ}$, and the $Y-$axis is oriented toward $l$ = 90$^{\circ}$ and $b=$ 0$^{\circ}$, and the disk rotates toward $l=$ 90$^{\circ}$; the velocity components are also oriented along these directions. In this convention, the Sun's orbital velocity vector are [U$_{\odot}$,V$_{\odot}$,W$_{\odot}$] = [$11.1$, $12.24$, $7.25$] km s$^{-1}$ \citep{Brunthaler2011}. The model has been rescaled to the Sun's Galactocentric distance, 8.3 kpc, and a local rotation velocity of $239$ km s$^{-1}$. For computation of the Galactic orbits, we have a simple Monte Carlo procedure and the Runge-Kutta algorithm of seventh-eight order elaborated by \citet{fehlberg68}. The uncertainties in the input data (e.g., distances, proper motions, and line-of-sight velocity errors listed in Table \ref{TableB1}) are assumed to follow a Gaussian distribution and were propagated as 1$\sigma$ variations in a Gaussian Monte Carlo re-sampling. We have sampled a half million orbits, computed backward in time for 3 Gyr. Fig. \ref{FigureC1} shows the probability densities of the resulting orbits projected on the equatorial (left column) and meridional (right column) Galactic planes, in the non-inertial reference frame where the bar is at rest. The orbital path (adopting central values) is shown by the black line in the same figure. The green and yellow colours correspond to more probable regions of the space, which are crossed more frequently by the simulated orbits.

We derive the distance to 2M12451043+1217401 using \texttt{StarHorse}\footnote{\url{ https://data. sdss.org/sas/dr14/apogee/vac/apogee-tgas/apogee_tgas-DR14.fits}}, a Bayesian distance estimator initially developed for APOGEE stars \citep{Queiroz2018}. For completeness, we provide dynamical solutions and orbit calculations for both the StarHorse distance of 13 kpc and the \citet{jones18} distance of 4.5 kpc. The two columns on the left of Fig. \ref{FigureC1} shows possible orbits given the \texttt{StarHorse} distance and different bar speeds (35, 40, 45 and 50 km s$^{-1}$ kpc), and the two columns on the right show the same figures for the \citet{jones18}  distance.

\begin{table}
	\setlength{\tabcolsep}{6.0mm}  
	\begin{tiny}
		\caption{Phase-space data}
		\label{TableB1}
		\begin{tabular}{ll}
			\hline
			\hline
			Coordinates & (J2000)     \\
			\,\,\,$(\alpha,\, \delta)$	& $(191^\circ .293486,\, 12^\circ .294486)$ \\
			\,\,\,$(l,\, b)$	& $(296^\circ.973729363,\, 75^\circ.0936328462)$\\  
			Heliocentric Distance    &  [kpc]  \\
			& $(4.54\pm 0.93)^{\rm a}$   \\
			& $(13.02\pm 1.49)^{\rm b}$    \\ 
			$\langle$RV$\rangle \pm \sigma_{\rm RV}$ & ${\rm \,[km\, s^{-1}}]$\\
			&(-82.8$\pm$9.9)$^{\rm c}$  \\
			Proper Motions   & $(\mu_\alpha\cos\delta\, ,\,\, \mu_\delta)$ \\
			& [${\rm mas\,yr^{-1}}$] \\
			& $(-1.32\pm0.12,-4.19\pm0.06)^{\rm d}$ \\
			\hline
			\hline
		\end{tabular} 
		\tablefoot{$^{a}$ Estimated distance computed by \citet{jones18}; $^{b}$ Estimated distance computed using the \texttt{StarHorse} code \citep{Queiroz2018}; $^{c}$ The average radial velocity of the binary system computed from the 24 visits of the APOGEE/DR14 spectra (see text); $^{d}$ Absolute proper motions from \citet{gaiadr2}.}
	\end{tiny}
\end{table}

\end{appendix}	
\end{document}